\documentclass[pra,aps,amsmath,showpacs,reprint,longbibliography,superscriptaddress]{revtex4-1}

\usepackage[dvips]{graphicx}
\usepackage[dvips]{color}
\usepackage{amsfonts}

\newcommand{\bra}[1]{\langle #1 | \,}
\newcommand{\ket}[1]{\, | #1 \rangle}
\newcommand{\braket}[2]{\langle #1 | #2 \rangle}

\newcommand{\mean}[1]{\langle #1 \rangle}
\newcommand{\om}{\omega}
\newcommand{\Om}{\Omega}

\newcommand{\De}{\Delta}
\newcommand{\eps}{\epsilon}
\newcommand{\taus}{\tau_\mathrm{s}}
\newcommand{\taut}{\tau_\mathrm{tr}}
\newcommand{\sig}{\sigma}
\newcommand{\deriv}[2]{\frac{\partial{#1}}{\partial{#2}}}
\newcommand{\sinc}{\mathrm{sinc}}

\begin{document}

\title{Optimizing inhomogeneous spin ensembles for quantum memory}

\author{Guy Bensky}
\affiliation{Department of Chemical Physics, Weizmann Institute of Science, 
Rehovot 76100, Israel}

\author{David Petrosyan}
\affiliation{Institute of Electronic Structure and Laser, 
FORTH, GR-71110 Heraklion, Crete, Greece}

\author{Johannes Majer}
\author{J\"org Schmiedmayer}
\affiliation{Vienna Center for Quantum Science and Technology (VCQ),
  Atominstitut, TU-Wien, Stadionalle 2, 1020, Vienna, Austria }

\author{Gershon Kurizki}
\affiliation{Department of Chemical Physics, Weizmann Institute of Science, 
Rehovot 76100, Israel}

\date{\today}

\begin{abstract}
  We propose a method to maximize the fidelity of quantum memory 
  implemented by a spectrally inhomogeneous spin ensemble. 
  The method is based on preselecting the optimal spectral portion 
  of the ensemble by judiciously designed pulses. 
  This leads to significant improvement of the transfer and storage
  of quantum information encoded in the microwave or optical field.
\end{abstract}

\pacs{03.67.Lx, 
42.50.Pq,  
76.30.Mi, 
85.25.Hv 
}

\maketitle

\section{Introduction}
Recent experimental demonstrations of strong coupling between spin
ensembles (SEs) and microwave photons of superconducting resonators
\cite{Schuster10,Wu10,Kubo10,Kubo11,Amsuss11,Bushev11} are an
important step towards realizing functional, hybrid quantum devices
\cite{Rabl06,Tordrup08A,Tordrup08L,Petrosyan08,Petrosyan09,Imamoglu09,Verdu09,Henschel10,Kasch10,Bensky11,Zhu11,Diniz11}.
Such hybrid devices may benefit from combining the advantageous
properties of very different subsystems, or ``blocks'': (i) a quantum
processor block containing, e.g., quantum dots \cite{Loss98,Hanson07}
or superconducting qubits \cite{Makhlin01,Devoret04,You05,Clarke08}
which can perform rapid quantum gate operations but are vulnerable to
decoherence due to their strong coupling to the environment and/or the
noise of the external controls; (ii) a quantum memory block consisting
of a SE of active dopants in a solid
\cite{Schuster10,Wu10,Kubo10,Kubo11,Amsuss11,Bushev11,Acosta09,Stanwix10}
or trapped ultracold atoms
\cite{Rabl06,Tordrup08A,Tordrup08L,Petrosyan08,Petrosyan09,Imamoglu09,Verdu09,Henschel10,Kasch10}
which are weakly coupled to the environment and therefore are suitable
for information storage; (iii) a quantum ``bus'' or interface, such as
a microwave cavity, whose interaction with the other blocks can be
quickly switched on and off by, e.g., tuning in and out of resonance
\cite{Blais04,Wallraff04,Sillanpaa07,Majer07}.  A related scenario
concerns reversible transfer and storage of optical excitations in SEs
used as memories for photonic quantum repeaters
\cite{Simon10,Lvovsky09,Afzelius09,Riedmatten08,Afzelius10,Moiseev10,Bonarota10}.

The potential advantages of hybrid quantum devices are countered by
decoherence during the transfer and storage of quantum information
(QI), which is rooted in the homogeneous (lifetime) broadening and the 
inhomogeneous spectral width of the SE constituting the memory block. 
Using magnetic dipole or optical Raman transitions can greatly prolong
the relaxation time of the spins. But since typically the coupling of
individual spins with the microwave cavity or optical field is weak,
QI must be collectively encoded in an ensemble of $N \gg 1$ spins,
which are prone to spectral broadening
\cite{Diniz11,Wesenberg11,Kurucz11}. This broadening introduces
dephasing on the memory block which can both hinder the transfer of QI
and limit its coherent storage time. Although conventional spin-echo
techniques \cite{Hahn50,Abragam61,AllEber1975,Tittel10} can prolong
the storage time, they may be impractical for spectrally broad SEs,
and they cannot improve the QI transfer (in fact they would preclude
the transfer altogether).

Here, after analyzing the fidelity of SE quantum memory, 
we present a method that can significantly improve its performance. 
The method is based on preselecting the optimal spectral portion 
of the inhomogeneously broadened ensemble using judiciously 
designed pulse(s), followed by the transfer of all the spurious 
spins to an auxiliary metastable state. Although the 
selected subensemble contains far fewer spins and therefore 
requires longer QI transfer times, we find that the resulting 
fidelity of the quantum memory during the QI transfer and storage 
is much higher under experimentally realistic conditions.

\section{Fidelity of an ensemble quantum memory} 

To set the stage for the discussion, assume that a qubit is encoded in
a SE as a superposition of its collective ground state $\ket{\psi_0}=
\ket{g_1,g_2,\ldots, g_N}$ and the fully-symmetrized single-excitation
(Dicke) state $\ket{\psi_1}= N^{-1/2} \sum_j \ket{j}$, where $\ket{j}=
\ket{g_1,g_2,\ldots, e_j,\ldots, g_N}$ denotes a state with only spin
$j$ excited.  If all the spins had the same resonant frequency
$\omega_0$ on the transition $\ket{g} \to \ket{e}$, the
single-excitation state would evolve in time $\tau > 0$ as
$\ket{\psi_1(\tau)}= N^{-1/2}\sum_j e^{-i \omega_0 \tau} \ket{j}$,
while the ground state $\ket{\psi_0}$ remains unchanged.  Yet, due to
inhomogeneous broadening, each excited spin has different resonant
frequency $\omega_j$. Then, even if the symmetric state $\ket{\psi_1}$
is prepared at $\tau = 0$, it would evolve into
$\ket{\tilde{\psi}_1(\tau)} = N^{-1/2}\sum_j e^{-i \omega_j \tau}
\ket{j}$.  We may thus define the storage fidelity at time $\tau \geq
0$ as the squared overlap of state $\ket{\psi_1(\tau)}$ with its
inhomogeneously-broadened counterpart $\ket{\tilde{\psi}_1(\tau)}$:
\begin{equation}
\label{eq-fidelitydef}
F (\tau) \equiv |\braket{\psi_1(\tau)}{\tilde{\psi}_1(\tau)}|^2
= \left| \frac{1}{N} \int n(\om) e^{-i (\om - \om_0) \tau} d \om \right|^2, 
\end{equation}
where $n(\omega)$ is the ensemble spectral density normalized to the total
number of spins, $\int n(\omega) d \omega = N$. Note that $F(0)=1$
for any $n(\omega)$. Since the ground state $\ket{\psi_0}$ does not
evolve in time, $F(\tau)$ quantifies how the fidelity of information 
\emph{already encoded} in a SE decreases over time. 
Specifically, for an ensemble with Lorentzian spectrum of width $\De$, 
$n(\omega)=n_0[1 + (\om -\om_0)^2/\De^2]^{-1}$, the fidelity loss is 
exponential in time, $F (\tau) = e^{-2\De \tau}$. As shown below,
Eq.~(\ref{eq-fidelitydef}) also characterizes the efficiency 
of excitation transfer to and from the SE.

\subsection{Transfer fidelity}
\label{sec-transfer}

Consider a single-mode field of a microwave cavity or 
an optical beam interacting with the SE. Assume that 
initially the field contains a single excitation (photon)
of frequency $\omega_0$ and the SE is in the ground state 
$\ket{\psi_0}$. Each spin $j$ in the ensemble interacts with
the field on the transition $\ket{g_j} \to \ket{e_j}$ with the 
coupling strength $\eta_j$. In the rotating wave approximation, 
neglecting the field and spin relaxations, their combined state 
at any time $\tau$ can be written as $\ket{\Psi(\tau)} = 
\alpha(\tau) \ket{1,\psi_0} 
+ \sum_j \beta_j(\tau) e^{-i (\om_j - \om_0) \tau} \ket{0,j}$,
where $\ket{1,\psi_0}$ refers to the state with a single photon
and all the spins in the ground state, and $\ket{0,j}$ denotes 
the field vacuum and the $j${th} spin excited. The probability
amplitudes $\alpha$ and $\beta_j$ evolve in time according to
\[
\begin{split}
\dot{\alpha} &= -i \sum_j \eta_j^* \beta_j e^{-i (\om_j - \om_0) \tau} , \\
\dot{\beta}_j &= -i \eta_j \alpha e^{i (\om_j - \om_0) \tau} , 
\end{split}
\]
which yields $\dot{\alpha}(\tau) = - \sum_j |\eta_j|^2 
\int_0^{\tau} d \tau' \alpha(\tau') e^{-i (\om_j - \om_0) (\tau - \tau')}$.
Assuming that the spin-field couplings $\eta_j$ are not correlated 
with the transition frequencies $\omega_j$ \cite{Kubo10,Kubo11,Amsuss11}, 
we finally obtain
\begin{equation}
\label{eq-transfer}
\dot{\alpha}(\tau) = - \bar{\eta}^2 N 
\int_0^{\tau} d\tau' \alpha(\tau')\sqrt{F (\tau-\tau')},
\end{equation}
where $\bar{\eta}^2 \equiv N^{-1} \sum_i |\eta_i|^2$ and $F(\tau)$ is
given by Eq.~(\ref{eq-fidelitydef}).  With $F(\tau)= 1$ for all
$\tau$, Eq.~(\ref{eq-transfer}) predicts Rabi oscillations between the
field and the SE according to $\alpha(\tau) = \cos (\bar{\eta}
\sqrt{N} \tau)$, so that at time $\taut \equiv \pi/\bar{\eta}\sqrt{N}$
there is a full retrieval of the excitation into the field,
$|\alpha(\taut)|^2 = 1$ [with the combined state $\ket{\Psi(\taut)} =
- \ket{1,\psi_0}$ acquiring a $\pi$ phase shift]. In the presence of
inhomogeneous broadening, however, $F(\tau)$ decreases with time,
resulting in damped Rabi oscillations.  The fidelity of the transfer
followed by retrieval is the value of $|\alpha(\taut)|^2$ after one
such Rabi oscillation.  For $F(\tau) \lesssim 1$ the fidelity loss is
$\mathcal{O} (1 - F)$.  Hence, increasing the storage fidelity
$F(\tau)$ for all times $\tau \in [0, \taut]$ also improves the
transfer fidelity.

\subsection{The optimal spectrum}
\label{sec-optimal-spectrum}

Our goal is to filter out of the SE with broad spectrum $n(\om)$ 
a subensemble with the spectrum $n'(\om)$ that will maximize the 
resulting fidelity $F'$ while still containing many spins 
$N' = \int n'(\om) d \om \gg 1$. Before discussing the filtering 
procedure, let us deduce the optimal spectrum of the subensemble. 
To this end, we may consider two different tasks: 
(\textit{a}) QI transfer to and from the memory, and 
(\textit{b}) QI storage in the memory for a specific time $\taus$.

(\textit{a}) For QI transfer, we require that the fidelity $F'(\tau)$
for the selected subensemble be high for all $\tau \in [0,\taut]$.
For any symmetric spectrum $n'(\om)$, Eq.~(\ref{eq-fidelitydef}) can
be expanded in a Taylor series
\begin{equation}
  \label{eq-fid-series}
  F'(\tau) = \left| \sum_{k=0}^\infty
  \frac{(-1)^k \tau^{2k}}{(2k)!} \mean{(\omega-\omega_0)^{2k}} \right|^2
  \approx 1-\mean{(\omega-\omega_0)^2} \tau^2,
\end{equation}
where $\mean{(\cdots)} \equiv \int n'(\om) (\cdots)d \om /\int n'(\om)
d \om$ denotes the average over the spectral distribution $n'(\om)$.
Hence, the spectral variance $\mean{(\omega-\omega_0)^2}$, which is
the leading term of the expansion, should be as small as possible for
a required number $N'$ of the selected spins. We then find that the
optimal spectrum is $n'(\om) = n(\om)$ for $|\omega-\omega_0| < \De$
and $n'(\om) = 0$ otherwise, where $\De$ is such that $N' =
\int_{\om_0 - \De}^{\om_0 + \De} n(\om) d \om$.  In other words, we
should select all the spins from the frequency interval $\om \in
[\om_0 - \De, \om_0 + \De]$, and none outside of it.  Assuming an
approximately constant original spectrum $n(\om) \simeq n_0$ around
$\om = \om_0$, we obtain $N' \simeq 2 \De n_0$ and
\begin{equation}
F'(\tau) \simeq \sinc^2(\De \tau) \simeq 1 - \mbox{$\frac{1}{3}$} \De^2 \tau^2
\quad (\De \tau \ll 1).
\label{eq-subensfidel}
\end{equation}

(\textit{b}) 
For QI storage over time $\taus$, we require that the fidelity 
$F'(\tau)$ be high at exactly $\tau = \taus$. For a symmetric 
spectrum $n'(\om)$, the fidelity of Eq.~(\ref{eq-fidelitydef}) reads
\begin{equation} 
F'(\taus) = \left|\frac{1}{N'} \int n'(\om) 
\cos([\om-\om_0] \taus) d \om \right|^2.
\end{equation}
Hence, we should select all the spins with frequencies $\om$ which 
maximize $\cos([\om-\om_0] \taus)$, i.e., a comb-like spectrum 
$n'(\om)$ peaked at $\om \simeq \om_k = \om_0 + 2 \pi k/\taus$,
where $k = 0, \pm 1, \pm 2, \ldots$ For a required number 
of selected spins $N'$, the optimal spectrum is then composed 
of a series of rectangular distributions around frequencies 
$\om_k$: $n'(\om) = n(\om)$ for $|\omega-\omega_k| < \De$ and 
$n'(\om) = 0$ otherwise (we note a similar result in \cite{Bonarota10}). 
Assuming the original spectrum changes little around each $\om = \om_k$, 
$n(\om) \simeq n_k$, we obtain $N' \simeq 2 \De \sum_k n_k$ and
\begin{equation}
F'(\taus)\approx \sinc^2(\De \taus) 
\simeq 1 - \mbox{$\frac{1}{3}$} \De^2 \taus^2 
\quad (\De \taus \ll 1) ,
\end{equation}
which has the same form as Eq.~(\ref{eq-subensfidel}), but with
important differences. First, the memory now rephases only at integer
multiples of $\taus$ \cite{Lvovsky09,Afzelius09,Bonarota10}. Second, 
the wider is the original spectrum $n(\omega)$, the larger number 
of peaks at $\omega_k$ with $n_k>0$ we can select. For a fixed number 
of spins $N'$, this allows for narrower $\De$, leading to higher fidelity 
at $\taus$.  Alternatively, for the same fidelity (fixed $\De$), this
results in a larger number of spins $N'$.

\section{Filtering of the ensemble}

We now present the filtering procedure in an ensemble of $N$ active 
dopants, while its optimization is the subject of the following Section.

\begin{figure}[t]
\includegraphics[width=0.6\linewidth]{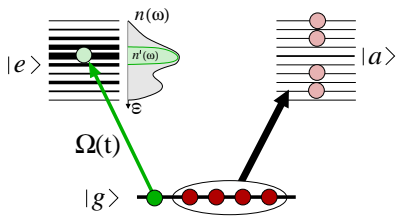}
\caption{Level scheme of dopants: $\ket{g}$ and $\ket{e}$ are 
the ground and excited states of the spin-$\frac{1}{2}$ subspace,
and $\ket{a}$ is the auxiliary metastable state. 
The subensemble with desired spectrum $n'(\om)$ is selected by the 
$\Om(t)$ field, which is followed by transferring all the remaining
dopants to $\ket{a}$.}  
\label{fig-3la}
\end{figure}

To select a subensemble of spins with spectrum $n'(\om)$, we employ 
another (auxiliary) long-lived state $\ket{a}$ outside the spin-$\frac{1}{2}$ 
subspace $\{\ket{g},\ket{e}\}$ of the dopants (see Fig.~\ref{fig-3la}). 
The preparation of the subensemble proceeds in three steps: 
(i) Starting with all the dopants in the ground state $\ket{g}$, 
apply an external pulse of Rabi frequency $\Om(t)$ that excites them 
to state $\ket{e}$. The duration $T$ of the pulse should be long enough, 
in order to select only the dopants with transition frequencies $\om$ 
within a range of $\De \sim 2 \pi/T$ around the desired frequency $\om_0$, 
while the shape of $\Om(t)$ is designed to optimize the resulting 
frequency spectrum.
(ii) Transfer all the dopants remaining in $\ket{g}$ to the auxiliary state 
$\ket{a}$ by another strong field, using an adiabatic sweep across the 
$\ket{g} \to \ket{a}$ transition \cite{Abragam61,AllEber1975}.
(iii) Return the dopants selected in step (i) from $\ket{e}$ to $\ket{g}$ 
by, e.g., the adiabatic transfer.

The chosen subensemble is now ready to use. Its spectrum is given by
$n'(\om) = n(\om) P(\om-\om_0)$, where  
\begin{equation}
P(\om - \om_0) = \left|\bra{e} T_+ \, e^{- i\int_0^T H(t) dt} \ket{g} \right|^2
\label{eq-selspectr}
\end{equation}
is the probability to excite the spin with transition frequency $\omega$ 
by the preparation pulse $\Omega(t)$ used in step (i). We consider only 
the amplitude-modulated field $\Om(t)$ with the fixed carrier frequency $\om_0$;
the corresponding Hamiltonian is $H(t) = \frac{1}{2} (\om - \om_0) \sig_z 
+ \Om(t) \sig_x$ with $\sig_{z,x}$ the Pauli spin operators. 
With the new spectrum $n'(\omega)$, the memory fidelity is given by
\begin{equation}
\label{eq-effective-fid}
\sqrt{F'(\tau)} = \frac{1}{N} \int n(\om) P(\om -\om_0) 
e^{-i (\om - \om_0) \tau} d \om .
\end{equation}

Ideally, we would like all the spins at the resonant frequency
$\om = \om_0$ to be selected, $P(0) = 1$, setting the pulse area
$A \equiv \int \Om(t)dt = \pi/2$ (a $\pi$-pulse). This still 
leaves us the freedom to choose the shape of $\Omega(t)$ so
as to maximize the fidelity in Eq.~(\ref{eq-effective-fid}). 

As an example, consider a square preparation pulse 
$\Om (t) = \pi/2T$ ($t \in [0,T]$), for which Eq.~(\ref{eq-selspectr}) 
can be be solved exactly, $P(\om - \om_0) = \frac{\pi^2}{4} 
\sinc^2 \big( \frac{1}{2} \sqrt{\pi^2 + (\om -\om_0)^2 T^2} \big)$. 
Using Eq.~(\ref{eq-fidelitydef}), we find linear fidelity loss 
$F'(\tau) \approx 1 - 4 \tau/ T$ for short times $\tau \ll T$.
With such a ``na\"ive'' choice of $\Om (t)$, the spectral variance 
$\mean{(\omega-\omega_0)^2}$ does not converge due to the long wings 
of $P(\om - \om_0)$ (see Fig.~\ref{fig-spectrum}), leading to poor 
fidelity (Fig.~\ref{fig-fidtrns}).

\section{Optimizing the preparation}

The ideal rectangular spectrum of subensemble found 
in Sec.~\ref{sec-optimal-spectrum} would require infinitely long 
preparation time. Our goal is therefore to find the optimal preparation 
pulse $\Om(t)$ of total duration $T$ which should be shorter than the 
spin relaxation time. To this end, we employ the method of Lagrange 
multipliers to maximize the fidelity $F'(\tau)$ for a given number 
of selected spins $N'$. [Recall that $N'$ determines the transfer 
time $\taut \propto (N')^{-1/2}$, which in turn should to be smaller 
than the decay time of the (cavity) field.] 
For convenience, we will actually maximize $N' \sqrt{F'}$ using 
the number spins $N'$ and the pulse area $A$ as the constraints. 
The resulting Euler-Lagrange equation reads
\begin{equation}
\deriv{\big[ N' \sqrt{F'(\tau)} \big]}{\Om(t)} 
= \lambda_1 \deriv{N'}{\Om(t)} + \lambda_2 \deriv{A}{\Om(t)} , 
\label{eq-eulerlagr}
\end{equation}
where $\lambda_1$ and $\lambda_2$ are the Lagrange multipliers.
We will pursue solutions of this equation yielding the optimal 
preparation pulse $\Om(t)$.

\subsection{Approximate analytic solutions}
\label{sec-approximate}

Although Eq.~(\ref{eq-eulerlagr}) can be studied numerically, it is 
instructive to solve an approximate version of this equation analytically.

From Eq.~(\ref{eq-selspectr}), in second order in $\Om$, we have 
\begin{equation}
P(\om) \approx \left| \int_0^T \Om (t) e^{- i (\om-\om_0) t } dt \right|^2 .
\label{eq-aprxselspectr}
\end{equation}
Assuming that $P(\om)$ is much narrower than the initial spectral 
distribution $n(\omega)$, we obtain
\begin{equation}
F'(\tau) \approx \frac{4 \pi^2 n_0^2}{{N'}^2} 
\left| \int_0^T \Om(t+\tau) \Om(t) dt \right|^2  ,
\label{eq-fidelity-prepared}
\end{equation}
with $N' \approx 2 \pi n_0 \int_0^T \Om^2(t) dt$ and $n_0 = n(\om_0)$.
Note that for $\tau > T$ the fidelity vanishes, i.e., one cannot 
store QI for time $\tau$ longer than the preparation time $T$.
Equation~(\ref{eq-eulerlagr}) now reduces to 
\begin{equation}
\Omega(t+\tau)+\Omega(t-\tau) = \lambda_1\Omega(t)+\lambda_2 ,
\label{eq-EL-sol}
\end{equation}
where we used the constraint $A = \int_0^T \Om(t)dt$. 

\begin{figure}[t]
\includegraphics[width=0.9\linewidth]{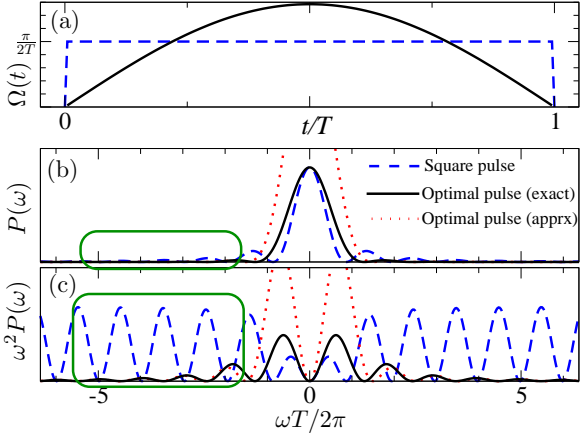}
\caption{(a) Square (blue dashed line) and optimal (black solid line)
  preparation pulses.  
  (b) The corresponding selection spectra $P(\om)$ calculated numerically.
  The approximate selection spectrum of Eq.~(\ref{eq-aprxselspectr}) 
  for the optimal pulse is also shown (red doted line); 
  the exact and approximate spectra differ significantly around resonance 
  but start to coincide for $\om T/2 \pi \gtrsim 1$.
  (c) Second moments $\om^2 P(\om)$ for the corresponding selection spectra.}
\label{fig-spectrum}
\end{figure}

\begin{figure}
\includegraphics[width=\linewidth]{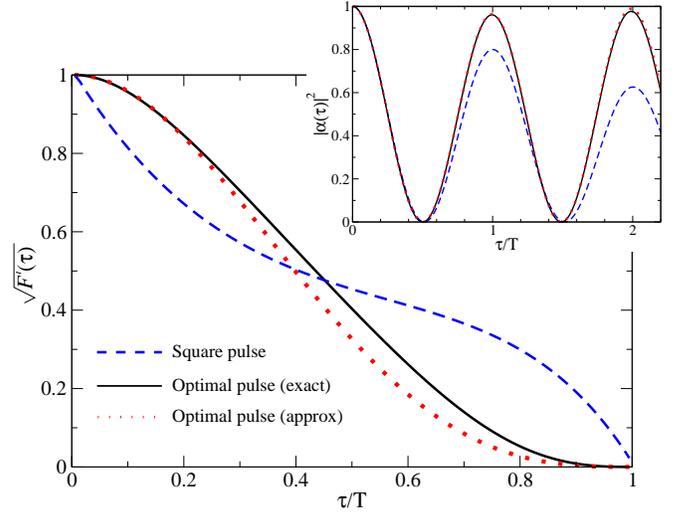}
\caption{Fidelity $F'(\tau)$ for subensemble selected by 
  square preparation pulse (blue dashed line) and by 
  optimal pulse of Eq.~(\ref{eq-optimal-solution-a}):
  exact solution (black solid line) and approximate analytical solution 
  of Eq.~(\ref{eq-optimal-fidelity-a}) (red dotted line).
  The duration of the pulses is $T = 10 \taut$.
  Inset: Rabi oscillations of a single excitation between the field 
  and the correspondingly selected subensemble.}
\label{fig-fidtrns} 
\end{figure}

Clearly, near the resonance $|\omega -\omega_0| \lesssim \max [\Omega(t)]$ 
Eq.~(\ref{eq-aprxselspectr}) is incorrect as the selection probability 
$P (\om \simeq \om_0) \lesssim \frac{\pi^2}{4}$ becomes larger 
than 1 [see Fig.~\ref{fig-spectrum}(b)]. This leads to the overestimate
of the number of selected spins $N'$, making the corresponding constraint
inexact. We will see below, however, that the maximal fidelity obtained 
under this approximation is close to the exact, numerically calculated 
fidelity, especially for $\tau \ll T$ [Fig.~\ref{fig-fidtrns}]. 
This is because the loss of fidelity is mainly due the wings of the 
selected spectrum, where the behavior of Eq.~(\ref{eq-aprxselspectr}) 
is correct. We also note that moderate nonuniformity of preparation 
pulse $\Om(t)$ for different spins would decrease $P (\om \simeq \om_0)$ 
and thereby the number of selected spins $N'$, but it would affect little 
the wings of the selected spectrum $P (\om)$ and the resulting fidelity $F'$.

We now seek the optimal preparation pulse which will maximize the fidelity 
(\textit{a}) over a continuous time interval $\tau \in [0,\taut]$ 
(for QI transfer), and 
(\textit{b}) at a specific time $\taus$ (for QI storage).

(\textit{a}) 
To maximize the fidelity for all times $\tau \leq \taut \ll T$, 
we notice that Eq.~(\ref{eq-EL-sol}) becomes independent of $\tau$: 
\begin{equation}
\ddot\Omega(t)=-\tilde\lambda_1\Omega(t)+\tilde\lambda_2 ,
\end{equation}
where $\tilde\lambda_1=-(\lambda_1-2)/\tau^2$ and
$\tilde\lambda_2=\lambda_2/\tau^2$ are the rescaled Lagrange multipliers.
The highest fidelity is then achieved with the pulse
\begin{equation}
\Om(t) = \Om_0 \sin(\pi t/T) , \label{eq-optimal-solution-a}
\end{equation}
where $\Om_0 = \frac{\pi^2}{4T}$, so that $A = \pi/2$ 
and $N' \approx \pi^5 n_0/16T$.
The optimal pulse (\ref{eq-optimal-solution-a}) and the corresponding
(exact and approximate) selection spectrum $P(\om)$ and its second
moment $\om^2 P(\om)$ are shown in Fig.~\ref{fig-spectrum}. Due to 
the suppressed wings of $P(\om)$, the spectral variance converges 
to $\mean{(\om -\om_0)^2} \sim \pi^2/T^2$, while it does not converge 
for the square preparation pulse. The resulting fidelity 
is shown in Fig.~\ref{fig-fidtrns} and is given by 
\begin{equation}
  F' (\tau) \simeq \left| \frac{(T-\tau) \cos ( \pi \tau/ T)} {T}
  + \frac{\sin ( \pi \tau/ T)} {\pi} \right|^2 
  \approx 1 - \frac{\pi^2\tau^2}{T^2} .
  \label{eq-optimal-fidelity-a}
\end{equation}
Hence, for short times $\tau \ll T$, the fidelity loss is quadratic in
$\tau$, which should be contrasted with linear fidelity loss for the
square preparation pulse.

In the inset of Fig.~\ref{fig-fidtrns}, we show the Rabi oscillations
of a single excitation between the field and the selected subensemble.
Using the fidelity of Eq.~(\ref{eq-optimal-fidelity-a}), we find 
the following solution of Eq.~(\ref{eq-transfer}) up to second order 
in $\taut/T$:
\begin{equation}
\alpha(\tau) \approx \left( 1- \frac{\taut^2} {T^2} \right)
\cos( \pi \tau/ \taut ) + \frac{\taut^2}{T^2} , 
\label{eq-optimal-transfer} 
\end{equation}
which yields $|\alpha(\taut)|^2 \approx 1 - 4 (\taut/T)^2$ 
and $|\alpha(2 \taut)|^2 \approx 1 + \mathcal{O} (\taut/T)^3$.
Remarkably, the probability of excitation retrieval into 
the field mode is higher at the end of the second oscillation 
at $\tau = 2\taut$ than the end of the first at $\tau = \taut$.
[Note that at the end of the second Rabi cycle the combined 
state of the field and SE has the initial phase,
$\ket{\Psi(2 \taut)} \simeq \ket{1,\psi_0}$.]
In general, for this spectrum the retrieval infidelity 
at even revivals is 3rd order in $\taut/T$, as opposed to 
2nd order at odd revivals. 

\begin{figure}[t]
\includegraphics[width=\linewidth]{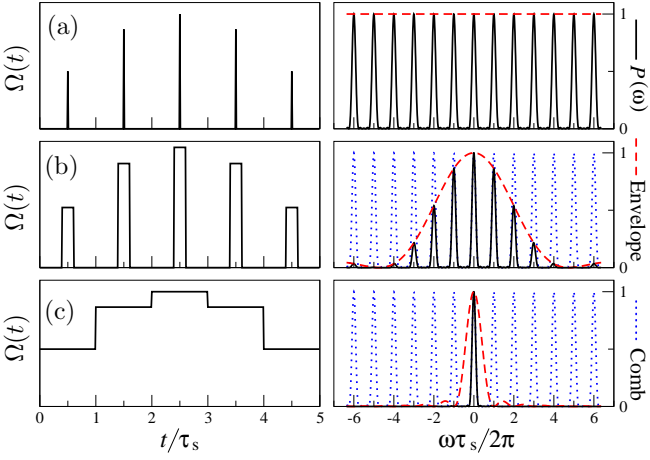}
\caption{Left column: 
  preparation fields $\Omega(t)$ of Eq.~(\ref{eq-optimal-solution-b}),
  optimized for storage time $\taus$, consisting of 
  (a) series of instantaneous pulses $\xi(t)=\delta(t-\taus/2)$, 
  (b) series of short pulses of width $\taus/5$, and 
  (c) step function constant during each $\taus$ interval $\xi(t)=1/\taus$. 
  The preparation time is $T = 5 \taus$.
  Right column:
  the resulting selection spectrum $P(\omega)$ of Eq.~(\ref{eq-selspctrtaus}) 
  given by the product of comb and envelope. 
  All the spectra have the same fidelity $F'(\taus)$ but different 
  number of spins $N' \propto \int_0^{\taus} \xi^2(t)dt$. }
\label{fig-comb}
\end{figure}

(\textit{b}) 
To maximize the fidelity at a specific storage time $\taus$,
we assume that the preparation time is a multiple of $\taus$,
$T = m \taus$ ($m \in \mathbb{Z}$). The solution of Eq.~(\ref{eq-EL-sol})
is then 
\[
\Om(t) = \Om_0 \, \xi(t - \taus \lfloor t/\taus \rfloor )
\, \sin \left(\pi \frac{ \lfloor t/ \taus \rfloor + 1}
{ m + 1} \right) , 
\]
where $\Omega_0 = \frac{\pi}{2} \tan\left(\frac{\pi/2}{m + 1}\right)$,
$\lfloor\cdots\rfloor$ denotes the integer part (floor) of the expression,
and $\xi(t)$ is a temporal profile within $\taus$ of unity area 
$\int_0^{\taus} \xi(t)dt=1$. Hence, the preparation pulse is
an $m$-fold repetition of $\xi(t)$, at each step $0 \leq n < m$ 
multiplied by a different amplitude,
\begin{equation}
\Omega(t) = \Omega_0 \, \xi(t - n \taus) \, 
\sin \left( \pi \frac{n+1}{m + 1} \right) \; \mathrm{for} \;   
 t \in [n \taus , (n+1) \taus) , 
\label{eq-optimal-solution-b}
\end{equation}
so that $A = \pi/2$ and 
$N' = \pi n_0 (m +1) \Omega_0^2 \int_0^{\taus} \xi^2(t)dt$.
The resulting approximate selection spectrum of Eq.~(\ref{eq-aprxselspectr}), 
\begin{equation}
P(\om) \approx \left|\sum_{n=0}^{m-1} e^{- i \omega n \tau_s} 
\sin \left(\pi \frac{n+1}{m+1} \right) \right|^2 |\tilde{\xi}(\om)|^2 ,
\label{eq-selspctrtaus}
\end{equation}
is a product of two terms: a ``comb'' term, which is a series of optimally 
shaped peaks spectrally separated by $2\pi/\taus$; and an ``envelope'' term,
$\tilde{\xi}(\om) \equiv \int_0^{\taus} e^{-i \om t} \xi(t) dt$, wider than 
$2\pi/\taus$. Examples of $P(\omega)$ for various $\xi(t)$ are shown 
in Fig.~\ref{fig-comb}. For a uniform $\xi(t) = 1/\taus$, $P(\om)$ reduces 
to a single peak, since the envelope has a width of $2\pi/\taus$, 
resulting in the minimal number of selected spins $N' \propto 1/\taus$. 
By contrast, choosing $\xi(t)$ to be a narrow pulse of width 
$\delta t < \taus$ yields multiple ($\sim \taus/\delta t$) peaks 
within the envelope, and hence larger number of selected spins, 
$N' \propto 1/\delta t$. A delta function, 
$\xi(t)=\delta(t-\tau/2)$, will give rise to infinitely many peaks 
and thus an infinite number of spins $N'\to\infty$. In practice, 
however, the number of selected peaks is limited by the width 
of the original SE spectrum $n(\omega)$. The resulting fidelity 
[Eq.~(\ref{eq-fidelity-prepared})] at time $\taus$ is now given by
\begin{equation}
F'(\taus) = \cos^2 \left(\frac{\pi}{T/\taus + 1} \right) .
\label{eq-specific-fidelity}
\end{equation}
Note that the choice of $\xi(t)$ does not affect the fidelity 
[unlike the number of selected spins $N' \propto \int_0^{\taus} \xi^2(t)dt$], 
which is to be expected, since different solutions must, by definition, 
yield the same maximal fidelity.

\subsection{Numeric solutions}

We now solve the Euler-Lagrange equation (\ref{eq-eulerlagr}) numerically 
using the exact selection spectrum of Eq.~(\ref{eq-selspectr}). 
We maximize the subensemble fidelity $F'(\taus)$ for two distinct 
situations:

\begin{figure}
\includegraphics[width=\linewidth]{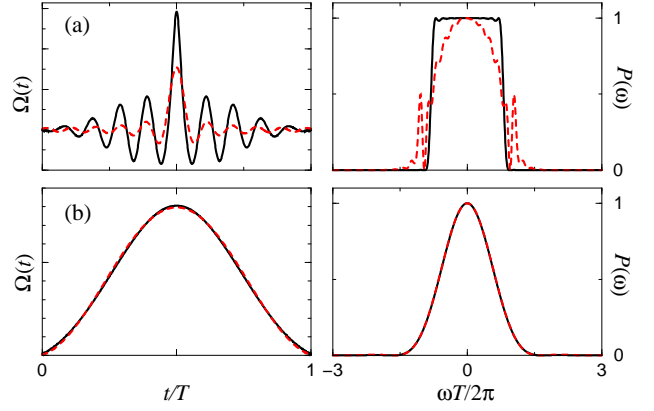}
\caption{Left column: preparation fields $\Om(t)$ optimized for 
  (a) long and (b) short preparation time $T$.
  In (a) the dashed (red) curve is the truncated $\sinc(t)$ function,
  and the solid (black) curve is the numerically optimized pulse.
  In (b) the dashed (red) curve is the $\sin(t)$ function 
  of Eq.~(\ref{eq-optimal-solution-a}) 
  and the solid (black) curve is the numerically optimized pulse.
  Right column: the resulting exact selection spectra $P(\omega)$.}
\label{fig-numsolv} 
\end{figure}

\paragraph{Stable spins --} In this scenario, we can use a long preparation 
time $T$ so as to achieve the optimal rectangular spectrum $P(\om)$ of the 
subensemble. Our main constraint is the number of selected spins $N'$. 
Clearly, for infinite $T$ the preparation pulse $\Om(t)$ is a $\sinc(t)$
function; for long but finite $T$, however, the optimal pulse is modified 
but still resembles the $\sinc(t)$, as shown in Fig.~\ref{fig-numsolv}(a). 

\paragraph{Short-lived spins --} Now the short lifetime of the spins 
severely limits the preparation time $T$. The resulting spectral width 
of the selected subensemble  $\De \sim 2 \pi /T$ will be wide enough 
to contain many spins. We can therefore disregard the constraint on $N'$,
focusing instead on achieving the narrowest possible selection 
spectrum  $P(\om)$ within a given preparation time $T$. We then find 
that the optimal pulse $\Om(t)$ is almost identical to that in 
Eq.~(\ref{eq-optimal-solution-a}), see Fig.~\ref{fig-numsolv}(b).

\subsection{Fidelity loss during the QI transfer and storage}

The preparation selects a subensemble of $N'\ll N$ spins 
with reduced spectral variance 
$\mean{(\omega-\omega_0)^2} \simeq \left( \frac{2}{\pi} \right)^8 (N'/n_0)^2$, 
but also results in a longer transfer time $\taut=\pi/\bar{\eta}\sqrt{N'}$.
During the transfer, the fidelity loss, or error due to the decay $\kappa$ 
of the (cavity) field is $\eps_\mathrm{tr} \sim \kappa \taut$, 
while the error accumulated during the storage time $\taus$ in the 
subensemble is $\eps_\mathrm{s} \sim \mean{(\omega-\omega_0)^2} \taus^2$.
Minimizing the total error $\eps = \eps_\mathrm{tr} + \eps_\mathrm{s}$
with respect to $N'$, we obtain 
\begin{equation}
\label{eq-error}
\min(\eps) \sim 2\left( \frac{\kappa^2 \taus}{\bar{\eta}^2 n_0} \right)^{2/5}
\;\; \mathrm{for} \;\; 
N' = \frac{\pi^{18/5}}{2^4} 
\left( \frac{\kappa n_0^2}{\bar{\eta} \taus^2} \right)^{2/5} .
\end{equation}
Hence, small error $\eps$ requires large cooperativity 
$\bar{\eta}^2 n_0 \gg \kappa^2 \taus$ of the field-subensemble coupling.

\section{Experimental considerations}

To illustrate the results of the foregoing discussion, we consider 
a SE of NV color centers in diamond coupled to a superconducting 
coplanar waveguide resonator \cite{Schuster10,Wu10,Kubo10,Kubo11,Amsuss11}.
The ground $\ket{g}$, excited $\ket{e}$ and auxiliary $\ket{a}$ 
states correspond, respectively, to the $m =0$, $m =1$ and $m =-1$ 
Zeeman sublevels of the ground electronic (spin-triplet) state
of the NV. Transitions $\ket{g} \to \ket{e},\ket{a}$ have 
the frequencies of around $2.88\:$GHz, and can be selectively addressed 
by the external $\sigma_{\pm}$-polarized microwave fields. 
In addition, a static magnetic field can be used to tune the 
transition $\ket{g} \to \ket{e}$ in and out of resonance with 
the cavity mode $\om_0$. The experimental inhomogeneous spectrum of 
the ensemble of $N \approx 10^{12}$ NV centers \cite{Amsuss11} 
has the total width of about $\Delta/2\pi \sim 7\:$MHz, 
composed of three partially overlapping Lorentzians of widths $\sim 2.6\:$MHz
split by $\sim 2.2\:$MHz due to the hyperfine coupling to the $I=1$ nuclear 
spin of the $^{14}$N atom. This results in a storage fidelity of 
$F(\taus) \approx 1 - \taus/60\,\mathrm{ns}$, while with the collective 
SE-cavity coupling strength $\bar{\eta} \sqrt{N} \simeq 2\pi \times 13\:$MHz 
\cite{Amsuss11}, the excitation transfer time is $\taut \approx 40\:$ns.

As an example, assume that one can achieve a cavity quality factor of
$Q=10^6$.  The photon lifetime in the cavity, $\kappa^{-1} \approx
55\:\mu$s, is then much longer than the transfer time $\taut$. This
allows a preparation that reduces the ensemble spectral width
$\Delta$, and the number of spins $N$, by a factor of $5 \cdot 10^{3}$
---almost $4$ orders of magnitude---while still keeping the transfer
time $\taut \approx 2.8\:\mu\mathrm{s} = 0.05\kappa^{-1} \ll
\kappa^{-1}$ well within the cavity lifetime.  The new subensemble,
created by the optimal preparation pulse $\Omega(t)$ of $T = 0.7\:$ms
duration, has the storage fidelity $F'(\taus) \approx 1 -
(\taus/0.22\,\mathrm{ms})^2$.  There is negligible loss of fidelity
during the QI transfer from the cavity field to the new ensemble,
which can now store QI for $\taus \simeq 50\:\mu$s with $95\%$
fidelity, compared to $3\:$ns with the original SE and $2.8\:\mu$s
inside the cavity.

Had we used the square preparation pulse of the same duration, the new
ensemble would have had a storage fidelity of $F'(\taus) \approx 1 -
\taus/0.17\,\mathrm{ms}$, which can store QI for $\taus \simeq 8.5
\:\mu$s with $95\%$ fidelity --- more than $5$ times worse than with
the optimal preparation.

\section{Summary and discussion}

Ensembles of long-lived two-level systems---spins---can serve 
as collective quantum memories, but their utility is often compromised
by the inhomogeneous spectral broadening, which can hamper both the
QI transfer to and from the SE and reduce the storage fidelity.
The fidelity is very sensitive to not only the width $\De$ but 
also to the profile of the inhomogeneous spectrum of the SE. 
Specifically, for a spectrum with long wings, such as a Lorentzian, 
the loss of fidelity during the storage time $\taus \ll \De^{-1}$
scales linearly with $\De \taus$, while for a spectrum with sharp 
cutoff at $\De$ the loss of fidelity is quadratic in $\De \taus$. 

In this paper, we have proposed and analyzed a method to select a
spectrally narrow subensemble of spins, which maximizes the fidelity
of QI storage and also improves the QI transfer from and to an
electromagnetic field. In our method, judiciously designed pulses of
finite duration---determined by the spin relaxation time---select the
optimal spectral portion of a large SE, while the remaining spins
making up the spurious part of the spectrum are discarded by
transferring them to an auxiliary metastable state. Our method is
applicable to microwave cavity (circuit QED) based hybrid quantum
systems involving QI processing qubits and SE quantum memories,
as well as to optical field storage in SEs. 
In the former case, the QI transfer time is limited by
the cavity field relaxation time, while in the latter case, it is the
interaction (transit) time of the optical pulse with the SE. Hence,
one always has to attain large cooperativity in ensemble-field
coupling by having many spins, and large optical depth.

Similar considerations may apply to the schemes involving noisy
processing qubits coupled directly to the spin-ensemble memories,
such as, e.g., the electron spin of a NV center interacting with 
the surrounding ensemble of long-lived nuclear spins
\cite{JelezkoPRL04,Dutt07,GaebelNAT06,NeumannSCI08,NeumannSCI10,BalasubramanianNAT09}.
Then in Eq.~(\ref{eq-error}) one would have to replace $\kappa$ by 
the qubit decoherence rate $\gamma$. As opposed to the cavity decay 
rate $\kappa$, which is difficult to change, $\gamma$ of a qubit 
can be suppressed by dynamical control methods
\cite{KofmanPRL01,GordonJPB07,GordonPRL08,ClausenPRL10,RaoPRA11,EscherJPB11},
which, for a fixed error $\eps$, will result in quadratic increase of the 
memory time $\taus \propto \eps^{5/2}/\gamma^2$.

\begin{acknowledgments}
We acknowledge the support of EC MIDAS STREP, DIP, 
the Humboldt-Meitner Award (G.K.), and the FWF Wittgenstein Prize (J.S.).
\end{acknowledgments}

\bibliography{qmemrefs}
\end{document}